\documentstyle[aps,prb,preprint,graphicx,floats]{revtex}

\tightenlines

\begin{document}

\draft
\title{Solution effects and the order of the helix-coil transition 
         in polyalanine} 
\author{Yong Peng  and Ulrich H.E. Hansmann } 
\address{Department of Physics, Michigan Technological University,
         Houghton, MI 49931-1291, USA}
\author{ Nelson A. Alves}
\address{Departamento de F\'{\i}sica e Matem\'atica, FFCLRP
     Universidade de S\~ao Paulo. Av. Bandeirantes 3900. 
     CEP 014040-901 \, Ribeir\~ao Preto, SP, Brazil}
\date{\today}
\maketitle
\begin{abstract}
We  study helix-coil transitions  in an all-atom model of
polyalanine. Molecules of up to length 30 residues are investigated 
by multicanonical simulations. Results from two  implicit solvent 
models are compared with each other and with that from simulations in  
gas phase. While  the helix-coil transition  is in all three models 
a true thermodynamic phase transition,  we find that its  strength 
is  reduced by the protein-solvent interaction term. The order of
the helix-coil transition depends on the details of the solvation 
term.   
\end{abstract}


\section{Introduction}
A key step in the folding of a protein  is the formation of 
secondary structure elements   such as $\alpha$-helices or
 $\beta$-sheets. In the case of $\alpha$-helices this process
resembles crystalization or melting and has been extensively
studied \cite{Poland}. In Refs.~\onlinecite{HO98c,AH99b,AH00b} evidence
was presented that this helix-coil transition is for 
polyalanine a true thermodynamic phase transition.
The latter result was obtained from gas-phase simulations  
where interactions between all atoms in the
molecule were taken into account. While  available data from 
gas-phase experiments \cite{Jarrold} support the numerical
results of these simulations, their relevance for
the biologically more important case of solvated molecules
needs to be established. First attempts in this direction can be found  
in Ref.~\onlinecite{MO1},  and in more detail in Ref.~\onlinecite{PH01h}
where  various solvent models are compared. 
These previous investigations are restricted to homopolymers of
chain lengths 10.  However, a detailed study on
how the thermodynamic  characteristics of the helix-coil 
transition change with the added solvent term  requires to go 
to larger chains and to probe the size dependence of that transition.
For this purpose, we have
performed multicanonical simulations of polyalanine chains
of length up to 30 residues. The protein-water interaction
is included by either a term that is proportional to the 
solvent accessible surface area  \cite{oons} or by a
distance-dependent permittivity  \cite{lavarty}. 
 Our results are compared with that of gas-phase simulations. We
find that the order of the helix-coil transition depends 
strongly on the details of the solvation term.


\section{Methods}
\noindent
Our investigation of the helix-coil transition for polyalanine is
based on a detailed, all-atom representation of that homopolymer. 
The interactions between the atoms are 
described by a standard force field, ECEPP/2  ({\bf E}mpirical 
{\bf C}onformational {\bf E}nergy {\bf P}rogram for {\bf P}eptides,
version 2) \cite{EC} as implemented in the program package SMMP 
({\bf S}imple {\bf M}olecular {\bf M}echanics for {\bf P}roteins)
 \cite{SMMP}:
\begin{eqnarray}
E_{ECEPP/2} & = & E_{C} + E_{LJ} + E_{HB} + E_{tor},\\
E_{C}  & = & \sum_{(i,j)} \frac{332q_i q_j}{\varepsilon r_{ij}},\\
E_{LJ} & = & \sum_{(i,j)} \left( \frac{A_{ij}}{r^{12}_{ij}}
                                - \frac{B_{ij}}{r^6_{ij}} \right),\\
E_{HB}  & = & \sum_{(i,j)} \left( \frac{C_{ij}}{r^{12}_{ij}}
                                - \frac{D_{ij}}{r^{10}_{ij}} \right),\\
E_{tor}& = & \sum_l U_l \left( 1 \pm \cos (n_l \chi_l ) \right).
\label{ECEPP/2}
\end{eqnarray}
Here, $r_{ij}$ (in \AA) is the distance between the atoms $i$ and $j$, and
$\chi_l$ is the $l$-th torsion angle.

The interactions between our homo-oligomer and water are approximated by
means of two implicit water models. In the first model 
one assumes that the free energy difference between  atomic groups 
immersed in the protein interior and groups exposed to  water   
is proportional to the solvent accessible surface area:
\begin{equation}
E_{solv}=\sum_i\sigma_iA_i,
\end{equation}
where $E_{solv}$ is the solvation energy, $A_i$ is the 
solvent accessible area (which depends on the configuration)
 of the $i$-th atom,  
and $\sigma_i$ is the  solvation parameter for the atom $i$.
For the present study we have chosen the  parameter set of 
Ref.~\onlinecite{oons}
that is often used in conjunction with the ECEPP force field. 
In the following, we will refer to this model as ASA (solvent
{{\bf A}ccessible {\bf S}urface {\bf A}rea).
Protein-water interactions are approximated differently 
in our second implicit water model to which we will refer as
DDE ({\bf D}istance {\bf D}ependent {\bf E}psilon). Here,  
a  distance  dependent  electrostatic permittivity  \cite{hingerty} 
\begin{equation}
\label{eps}
\varepsilon(r) = D-\frac{D-2}{2}[(sr)^2+2sr+2]e^{-sr} \, ,
\end{equation}
is introduced to model electrostatic interaction between the 
protein atoms in the presence of water.  The parameters $D$ and $s$  
are chosen such  that for large distances the permittivity takes 
the value $\varepsilon \approx 80$ of  bulk water, and 
 $\varepsilon=2$ for short distances (protein interior space). 

In detailed models of biological macromolecules the various
competing interactions lead to an energy landscape with a 
multitude of local minima separated by high energy barriers. 
Canonical Monte Carlo or molecular dynamics
simulations likely will  get trapped in one of these
minima and not thermalize within the available CPU time. Only recently,
with the introduction of new and sophisticated algorithms such as
{\it generalized-ensemble} techniques \cite{Review} was it
possible to alleviate this problem in  protein simulations \cite{HO}.
For this reason, we rely  in our project  on one of these 
sophisticated techniques, multicanonical sampling \cite{MU}, whose
performance for polyalanine is extensively discussed in 
Ref.~\onlinecite{OH95b}.

The multicanonical algorithm \cite{MU}
assigns a weight $w_{mu} (E)\propto 1/n(E)$ to conformations with energy $E$.
Here, $n(E)$ is the density of states.
A  simulation with this weight
will  lead to a uniform distribution of energy:
\begin{equation}
  P_{mu}(E) \,  \propto \,  n(E)~w_{mu}(E) = {\rm const}~.
\label{eqmu}
\end{equation}
Thus, the simulation generates a 1D random walk in the
energy space,
allowing itself to escape from any  local minimum.
Since a large range of energies is sampled, one can
use the reweighting techniques \cite{FS} to  calculate thermodynamic
quantities over a wide range of temperatures $T$ by
\begin{equation}
<{\cal{A}}>_T ~=~ \frac{{\int dx~{\cal{A}}(x)~w^{-1}(E(x))~
                 e^{-\beta E(x)}}}
              {{\int dx~w^{-1}(E(x))~e^{-\beta E(x)}}}~,
\label{eqrw}
\end{equation}
where $x$ stands for configurations 
and $\beta$ for the inverse temperature, $\beta = 1/k_B T$.
Note that the weights are
not {\it a priori} known in multicanonical simulations  and
 estimators  need to be determined. This is often done 
by an iterative procedure described in detail in 
Refs.~\onlinecite{Berg,PH01h}, with which we needed  between 
100,000 ($N=10$) and 800,000 ($N=30$)
sweeps for the weight factor calculations. All thermodynamic quantities 
are  estimated then from one production run of $N_{sw}$ Monte Carlo sweeps 
starting from a random initial conformation. 
 Our  emphasis is on the ASA solvent for which we have 
chosen  $N_{sw}= 6,000,000$, while for the DDE solvent approximation  
our simulations rely on $N_{sw}= 2,000,000$ sweeps. In all runs, we 
store every 10th sweep for further analysis. Our error bars are obtained 
by the jackknife method, with 12 bins for ASA, and 8 bins for the 
DDE model. The results of these implicit solvent
simulations  are compared with that of polyalanine
in gas phase \cite{HO98c,AH99b,AFH01d} that rely on
 $N_{sw}$=500,000, 500,000, 1,000,000, and 3,000,000 sweeps
for $N=10$, 15, 20, and 30, respectively.

Thermodynamic quantities that we  calculate from these 
multicanonical simulations for our three models (gas phase,
ASA and DDE) include  the average energy, 
specific heat, helicity and susceptibility. We further
evaluate the complex partition function zeros whose analysis 
was introduced by us recently as a tool in studies of structural 
transitions in biomolecules \cite{AH99b,AH00b,AFH01d}.
Such an analysis is possible because the multicanonical algorithm 
allows one to calculate reliable estimates for the spectral density:
\begin{equation}
  n(E) = P_{mu} (E) w^{-1}_{mu} (E)~.
\end{equation}
We can therefore construct
the partition function in the complex variable
$u$ from these estimates,
\begin{equation}
     Z(\beta) = \sum_{E} n(E) u^{E} ,                         \label{eq:r1}
\end{equation}
where $u=e^{-\beta}$.
The complex solutions of the partition function (the so-called
Fisher zeros \cite{fisher,itzykson}) correspond to the complex
extension of the temperature variable and determine the critical 
behavior of the model.  Partition function zeros analysis is used here
again to study for polyalanine the effects of the two implicit
solvent models on the characteristics of the helix-coil transition.


\section{Results and Discussion}
We could show in previous work that polyalanine exhibits a pronounced 
helix-coil transition in gas phase \cite{OH95b,HO98c,AH99b,AH00b}.
In this paper, we investigate now how the characteristics of that
transition change in the presence of an implicit solvent.
A natural order parameter 
for the helix-coil transition is $q_H =<n_H(T)>/(N-2) $, i.e. the
average number of helical residues divided by the number of residues
that can be part of an $\alpha$-helix.  A residue is considered 
as ``helical'' if its backbone dihedral angles $(\phi,\psi)$ take 
values in the range $(-70^{\circ}\pm 30^{\circ},-37^{\circ}\pm30^{\circ})$ 
 \cite{OH95b}. The normalization factor $N -2$ is chosen instead
of $N$, the number of residues,  because the terminal residues are 
flexible and are usually not part of an $\alpha$-helix.  We  start our 
analysis by displaying in Fig.~1a and 1b over a larger temperature range
this order parameter  as calculated from ASA (Fig.~1a) and DDE (Fig.~1b)
simulations. We notice in both figures a clear separation 
between a high-temperature phase with few helical residues and 
a low-temperature phase that is characterized by a 
single $\alpha$-helix. The free energy difference per residue  
 $\Delta g_{hc}(T) = (G_{helix} (T) - G_{coil} (T))/(N-2)$  
between helix and coil configurations is plotted as a function 
of temperature in the inlets. At high temperatures, $\Delta g_{hc}$ 
is positive and coil configurations are favored. On the other hand, below
a transition temperature $T_c$ (that depends on the chain length) helical
states are favored and the  free energy difference is consequently negative.

The  reason for the stability of the helical state at low temperature
is the large difference in intramolecular energy to  the coil structures
(that have a much larger entropy and dominate in the high temperature
phase).  For instance, at $T= 275$ K, well into the low-temperature
phase, we find in ASA simulations for polyalanine chains of 
length $N=30$ an average potential energy difference
\mbox{$<\Delta E_{hc}(ASA)>$} 
 $= <E_{tot}({helix}) - E_{tot} ({coil})> = -43.5(2.9)$ kcal/mol. 
This energy difference is the sum of 
two competing terms. Helices are favored over coil configurations 
by an ECEPP/2 energy difference of 
$\Delta E_{ECEPP/2} = -55.2(3.7)$ kcal/mol, 
however, the ASA solvation term
favors coil configurations and decreases that term 
by $\Delta E_{ASA} = 9.2(1.9)$  kcal/mol. In DDE simulations,
we have only one energy term, and here we find 
\mbox{$<\Delta E_{hc}(DDE)>$} $ = -73.0(1.2)$ kcal/mol. 
 For comparison, we have found in previous gas-phase simulations 
of polyalanine chains of the same length at this temperature
a value of \mbox{$<\Delta E_{hc}>$} $=-75.9(1.8)$ kcal/mol.
Hence, in both solvent models, helices are energetically less favored
than in gas-phase. Consequentely, the free energy difference per residue
between helix and coil configurations is,  with
$\Delta g_{hc}({ASA})= -0.60$ kcal/mol and 
$\Delta g_{hc}({DDE}) = -1.07$ kcal/mol,
smaller than in gas phase simulations where we have
 $\Delta g_{hc}({gas}) = -1.17$  kcal/mol.
This  result is in contradiction with  work 
by Mortenson {\it et al.} \cite{Wales} who claim that 
solvent effects enhance helix formation, but in agreement
with other recent studies by Vila {\it et al.}\cite{VRS00} and
Levy {\it et al.}\cite{LJB01}.

In  simulations with both solvent terms, the free energy 
difference increases with the size of the molecule  making 
the transition between both states sharper as the system size 
increases (see the inlets of Fig.~1a and 1b). This indicates 
that we observe  in both both solvent models
a phase transition for polyalanine.  Such a phase transition 
requires long range order in the infinite system. We test
our model for this criterion by calculating
the helix propagation parameter $s$ and the nucleation 
parameter $\sigma$  of the Zimm-Bragg model \cite{ZB}. These
two quantities are related to the average number
of helical residues $<n>$ and the average length $<\ell>$ of a helical
segment. For large number $N$ of residues, we have
\begin{eqnarray}
{{<n>} \over N}~ &=& ~{1 \over 2} - {{1-s} 
       \over {2\sqrt{(1-s)^2 + 4s \sigma}}}~, \\
<\ell>~~ &=& ~1 + {2s \over {1-s+\sqrt{(1-s)^2 +4s \sigma}}}~.
\end{eqnarray}

Table 1 lists the  values  of $s$ and $\sigma$ as calculated by
this equation for all chain lengths and both models. For comparison,
we have added also the corresponding values for polyalanine in gas
phase as listed in Ref.~\onlinecite{HO98c}.
We are  especially interested in the nucleation parameter $\sigma$
that characterizes the probability of a helix segment to break
  apart into two pieces. In Fig.~2, we draw a log-log plot of this
quantity as a function of chain length for $T=275$ K, deep in the low
temperature region. For both  implicit solvent models the data points 
form a straight line. Hence, $\sigma$ can be described by a power-law: 
\begin{equation}
<\sigma(N)> = \sigma_0 N^{-c}~,
\end{equation}
with $\sigma^{ASA}_0 = 0.6(4) $,  $c^{OONS} = 1.2(1)$ and 
 $\sigma^{DDE}_0 = 0.8(2)$, $c^{DDE}= 1.2(1)$.
It follows that $<\sigma (\infty)> = 0$ for both  solvent models, 
i.e.  the probability for a helical segment to break into pieces approaches
in both models to zero for infinite chain length. The same result was found
in Ref.~\onlinecite{HO98c} for polyalanine in gas phase, and the corresponding
values of $\sigma$ are  shown in Fig.~2 for comparison. 
It follows that polyalanine exhibits  long-range order in the 
low-temperature phase (and therefore allows for a phase transition) 
independently on whether the polymer is
simulated in gas phase or with ASA or DDE  implicit solvent.
We remark that our $s$-parameter values  (see table 1) are in 
agreement with the experimental results of Ref.~\onlinecite{CB} where they 
list values of $s$(Ala) between $1.5$ and $2.19$.

In order to analyze the helix-coil transition in more detail, and to
compare our results with that of previous gas phase simulations, we
determine first the transition temperatures from the corresponding 
plots in the specific heat $C_N(T)$ that are shown in Fig.~3a (ASA) and 
3b (DDE). The so obtained values are listed in table 2 together with
 the corresponding gas-phase values of Refs.~\onlinecite{HO98c,AH99b}.
As already observed in Ref.~\onlinecite{PH01h}, the  transition 
temperatures are for ASA simulations lower than in gas phase, 
but higher for DDE simulations. However, the differences decrease 
with chain length.  If we  extrapolate the listed temperatures 
to the  infinite chain limit by
$T_c (L) = T_c(\infty) - a\, e^{-bN}$, 
we find for ASA simulations as the critical temperature
$T_c(\infty) = 480$ K,  which is only 30 K
lower than the corresponding value for gas-phase simulations:
 $T_c (\infty) = 514$ K. For DDE simulations  we find $T_c(\infty) = 525$ K,
i.e. the difference to gas-phase  is only $\approx 10$ K.
We note that the transition temperatures are outside of the
range of physiologically relevant temperatures indicating  limitations
of our energy functions in protein folding studies.

Further information on the helix-coil transition can be obtained from 
the finite-size scaling analysis of the peak of the specific heat. 
We show in Fig.~4 a log-log plot of the maximum
value $C_N^{\rm max}$ of the specific heat as calculated from gas-phase,
ASA and DDE simulations for all four chain lengths. One expects that
for sufficiently large chains  $C_N^{\rm max}$  can be described by the
scaling relation:
\begin{equation}
C_N^{\rm max} \propto  N^{\alpha/d\nu}~.
\label{alpha}
\end{equation}
Only the gas-phase values fall in the log-log plot  on a straight line. 
This allowed us  to calculate in Ref.~\onlinecite{HO98c,AH99b} the specific
heat exponent $\alpha$ of polyalanine in gas phase: $\alpha = 0.86(10)$.  
However, in  simulations with one of solvent approximations,  $C_N^{\rm max}$  
can not be described by a straight line in a log-log plot. They rather 
seem to converge towards a constant value. Such a behavior is expected 
if the system has a specific heat exponent $\alpha = 0$, in which case 
logarithmic corrections need to be taken into account when describing 
the scaling of the specific heat maximum.  A similar picture (Fig.~5) 
appears for the scaling of the peak in the susceptibility,
\begin{equation}
\chi(T) = (<n_H^2> - <n_H>^2)/(N-2) \, ,
\end{equation}
which is expected to follow the scaling law,
\begin{equation}
\chi_N^{\rm max} \propto  N^{\gamma/d\nu}~.
\label{gamma}
\end{equation}
Fig.~6 displays the corresponding log-log plot of this quantity for
all three models and the four chain lengths.  Again, the gas-phase values
lay on a straight line and allow us to calculate an estimate for the 
susceptibility exponent $\gamma = 1.06(14)$ \cite{HO98c}.  However, 
for ASA and DDE simulations, the peak values  converge again towards
a constant value indicating $\gamma = 0$ and logarithmic corrections to 
scaling.

Calculation of these correction terms is in general
difficult and requires very large chain lengths and 
statistics (see, for instance, Ref.~\onlinecite{ADH97} where
the problem was tackled for the $2D$-Ising model).
For polyalanine, the available molecule sizes and statistics 
do not allow  for a reliable fit where the logarithmic 
correction terms are taken into account.  Instead, we pursue  
a different way.  We have shown in recent articles 
  \cite{AH99b,AH00b,AFH01d,AHP01e} 
that considerable information on structural transitions in 
biological molecules can be gained from an analysis of the
partition function zeros of the system.  We display 
in tables III, IV and V, respectively, the first  complex 
zeros for polyalanine chains in the gas phase, and 
for both  solvent models. Methods and the evaluation
of these zeros are discussed in Ref. \onlinecite{AFH01d} 
for the gas-phase model. 
 We note that  constructing the partition function is more difficult
for the solvated molecules. Even while the larger number of sweeps
is larger than in gas phase we can obtain only the first three zeros 
and not  four zeros as was possible for the gas phase model. 

One way of extracting  information on the strength of
transitions in molecules from
the partition function zeros is the approach by Janke and Kenna \cite{JK} 
who assume that the zeros condense for large systems
on a single line,
\begin{equation}
u_j = u_c + r_j\,{\rm exp} (i \varphi)~,                   \label{a1}
\end{equation}
with the cumulative density of zeros given by the
average
\begin{equation}
  G_N(r_j) = \frac{2j-1}{2N}  \, .                      \label{poly1}
\end{equation}
Here, $j$ labels the complex zeros in order of increasing imaginary part,
 $u_j = {\rm exp}(-\beta_j), j=1,2, \cdots $. Janke and Kenna postulate
 that this density scales for sufficiently large chains as \cite{JK}:
\begin{equation}
\frac{2j -1 }{2N} = a_1 ({\rm Im}\, u_j(L))^{a_2} + a_3 \,.\label{a5}
\end{equation}
A necessary condition for the existence of a phase transition
is that $a_3$ is compatible with zero, else it  indicates
that the system is in a well-defined phase.  Indeed, we obtain from
tables IV and V values $a_3$ that are compatible with zero for
all chain lengths. The values of
the constants $a_1$ and $a_2$ then  characterize the phase transition.
For instance, for first order transitions
the constant $a_2$  should take values $a_2 \sim 1$,
 and in this case the slope of this equation is
related to the latent heat\cite{JK}.
On the other hand, a value of $a_2$ larger than $1$
indicates a second order transition whose specific heat
exponent is given by $\alpha = 2 -a_2$.

Table VI lists the parameter $a_2(N)$ of Eq.~(\ref{a5}).
The estimates are less precise for the implicit solvent models
than for polyalanine in gas phase. However, the values clearly 
exclude the possibility of a first order transition for the 
solvated molecule while  for the polymer in gas phase we obtained 
as our best value for $a_2 = 1.16(1)$ but could not exclude
a value $a_2 = 1$ when we fit all chain lenghts \cite{AFH01d,AHP01e}
 (i.e. the possibility of a
first order phase transition). On the other hand, 
a multiple fit\cite{JK} in the parameters $j$ and $N$   leads 
in the case of  the ASA  solvent  to $a_2 = 1.79(8)$ (all zeros considered)
and  $a_2 = 1.90(9)$  if only the lowest eight zeros are considered.
Similarly, one finds for DDE solvent $a_2 = 1.74(11)$ (all zeros 
considered) and $a_2 = 1.81(24)$  if one  considers only the lowest 8 
zeros. The corresponding estimates for the specific heat 
exponent, $\alpha = 0.10(9)$ for ASA solvent and $\alpha = 0.19(24)$ 
for DDE solvent, are  small and within the errorbars compatible
with zero. Hence, they support our conjecture that 
$\alpha = 0$ for polyalanine with an implicit solvent. Note, that
this calculations demonstrates the advantages of partition function zero
analysis over other approaches:  our data do not allow us to
obtain a reliable estimate for $\alpha$ for either of the two solvent models 
from the  finite-size scaling of the peak in the specific heat 
(see Fig.~4) while this is possible with partition function zero analysis.

Our analysis of the maxima of the specific heats and the partition 
function zeros indicate that the helix-coil transition in 
polyalanine with a DDE or ASA solvent  is second order with 
exponents that are consistent with $\alpha =  \gamma =0$, and
by means of the hyperscaling relation $\alpha = 2 - d\nu$ with
 $d\nu = 2$.
Hence, the order of the
transition in the solvated molecule is fundamentally different from the
one in gas phase whose exponents $\alpha = 0.86(14), \gamma = 1.06(10)$
and $d\nu = 0.93(7)$ \cite{HO98c,AH99b} are consistent with a 
(weak) first order transition or a strong second order transition.

The weakening of the helix-coil transition in models that
try to account for protein-water interactions is not unexpected.
A large part of the energy gain through helix formation comes 
from the formation of hydrogen bonds between a residue and the 
forth following one that characterizes an $\alpha$-helix. 
Within water  this process competes with the entropically more 
favorable formation of hydrogen bonds with the surrounding water. 
Another factor are hydrophobic forces that penalize the extended 
structure of an $\alpha$-helix for the hydrophobic alanine. The
corresponding smaller gain in potential energy through helix 
formation in water when compared to gas phase is  observed by
us for both implicit solvent models. Hence, one can expect that 
helix-formation is more favored
in gas-phase than in solvent. This is consistent with our observation
of a weaker transition in the solvated molecule than observed in
gas phase.


\section{conclusion}
We have studied the helix-coil transition in two models of polyalanine
that attempt to approximate the interaction of the molecule with the
surrounding water by an implicit solvent. We find that the helix-coil 
transition in these models is a true thermodynamic phase transition. 
However, while we have for polyalanine in gas phase
critical exponents that are consistent with  a (weak) first order phase 
transition, our  results 
rather indicate for ASA and DDE solvent models
 a (weak) second order transition with critical exponents
$\alpha$ and $\gamma$ close to zero  and $d\nu$ close to two. This change 
in the strength of the helix-coil
transition is consistent with our understanding of the protein-water
interactions. While our results  show that the two implicit solvation
models  describe  qualitatively correct the physics of protein-water
interaction, the un-physiologically high transition temperatures
demonstrate also the limitations of these models.
\\
 
\noindent
{\bf Acknowledgements}: \\
 N.A. Alves gratefully acknowledges support by CNPq (Brazil), and
 U.H. Hansmann support by a research grant
from the National Science Foundation (CHE-9981874).
 Part of this work was done while
U.H. visited the Ribeir\~ao Preto campus of the University of S\~ao Paulo.
He thanks FAPESP for a generous travel grant and the Department of
Physics and Mathematics for kind hospitality.
 The authors are also grateful to DFMA (IFUSP) for the computer facilities
made available to them.

\clearpage

 

\vfil

\clearpage 
\newpage

\begin{table}[ht]
\renewcommand{\tablename}{Table}
\caption{Nucleation parameter $\sigma$ and helix-propagation parameter
 $s$ of the Zimm-Bragg model, as calculated from multicanonical simulations
 for polyalanine chains of length $N=10,15,20$ and $30$ in DDE or ASA
 solvent. The gas phase results (GP) are taken from Ref.~2}
\begin{center} 
\begin{tabular}{lccccccccc} \\
\\[-0.3cm]
 &&\multicolumn{2}{c}{\rm GP} && \multicolumn{2}{c}{\rm ASA}
 &&\multicolumn{2}{c}{\rm DDE}\\
 $N$&& $\sigma$ & $s$       & & $\sigma$ &  $s$     & & $\sigma$ &$s$ \\
\\[-0.35cm]
\hline
\\[-0.3cm]
 10 && 0.126(4) & 1.561(28) & & 0.122(1) & 1.345(29)& & 0.130(1) & 1.767(6)\\  
 15 && 0.074(2) & 1.679(10) & & 0.064(1) & 1.501(13)& & 0.073(1) & 1.860(11)\\
 20 && 0.056(1) & 1.780(13) & & 0.045(1) & 1.561(12)& & 0.052(1) & 1.890(10)\\
 30 && 0.036(1) & 1.845(25) & & 0.032(1) & 1.501(96)& & 0.034(1) & 1.934(11)\\
\end{tabular}
\end{center} 
\end{table}
\begin{table}[ht]
\renewcommand{\tablename}{Table}
\caption{\baselineskip=0.8cm Numerical results for various polyalanine 
       chain lenghts $N$: critical temperature $T_c$ 
       defined by the maximum of specific heat $C^{\rm max}$ and the
       maximum of susceptibility $\chi^{\rm max}$. The gas phase results (GP)
       are taken from Ref.~2.}
\begin{center}
\begin{tabular}{lcrclcrclcrcl}\\
\\[-0.3cm]
     & &    \multicolumn{3}{c}{\rm~~GP} & &
            \multicolumn{3}{c}{\rm ASA} & &
            \multicolumn{3}{c}{\rm ~DDE} \\
$~N$ & &$T_c$~~ &~$C^{\rm max}$ &$\chi^{\rm max}$ & &
        $T_c$~~ &~$C^{\rm max}$ &$\chi^{\rm max}$ & &
        $T_c$~~ &~$C^{\rm max}$ &$\chi^{\rm max}$ \\
\\[-0.35cm]
\hline
\\[-0.3cm]
~10  & & 427(7)   &~~8.9(3)        & 0.49(2)          & &
         333(2)   & 10.2(1)        & 0.613(7)         & &
         482(1)   &~~9.9(1)        & 0.605(5)  \\
~15  & & 492(5)   & 12.3(4)        & 0.72(3)          & &
         403(3)   & 13.1(2)        & 0.81(1)          & &
         517(3)   & 12.3(2)        & 0.71(1)  \\
~20  & & 508(5)   & 16.0(8)        & 1.08(3)          & &
         430(2)   & 14.5(4)        & 0.91(3)          & &
         518(4)   & 14.0(4)        & 0.79(2)    \\
~30  & & 518(7)   & 22.8(1.2)      & 1.50(8)          & &
         461(3)   & 15.5(1.1)      & 1.00(8)          & &
         523(7)   & 13.9(6)        & 0.81(4)     \\
\end{tabular}
\end{center}
\end{table}


\begin{table}[ht]
\renewcommand{\tablename}{Table}
\caption{\baselineskip=0.8cm Partition function zeros for polyalanine
          in the gas phase. The data are taken from 
          Ref.~19.}
\begin{center}
\begin{tabular}{lcccccccc}\\
\\[-0.3cm]
$~N$~~&Re$(u_1)$ &Im$(u_1)$    & Re$(u_2)$ & Im$(u_2)$ &Re$(u_3)$ &Im$(u_3)$ 
      &Re$(u_4)$ &Im$(u_4)$  \\
\hline
\\[-0.3cm]
~10 &0.30530(12) &0.07720(14)  &0.2823(13) &0.13820(61)&0.2459(72)&0.1851(63)
    &0.172(11)   &0.2200(71)    \\
~15 &0.356863(61)&~0.053346(39)&0.34167(60)&0.10440(59)&0.3331(48)&0.1454(28)
    &~0.3067(81) &0.1689(32)    \\
~20 &0.374016(41)&~0.042331(45)&0.36161(27)&0.08109(24)&0.3569(27)&0.1154(13)
    &~0.3336(56) &0.1470(27)    \\
~30 &0.378189(19)&~0.027167(32)&~0.37399(14)&~0.05420(27)&0.3693(11)
    &0.0804(13)  &~0.35854(63) &0.1022(43)    
\end{tabular}
\end{center}
\end{table}


\clearpage 
\newpage

\begin{table}[ht]
\renewcommand{\tablename}{Table}
\caption{\baselineskip=0.8cm Partition function zeros for polyalanine
 chain lenghts $N$ with ASA solvent.}
\begin{center}
\begin{tabular}{lcccccc}\\
\\[-0.3cm]
$~N$~~&Re$(u_1)$ &Im$(u_1)$  & Re$(u_2)$ & Im$(u_2)$ &Re$(u_3)$  &Im$(u_3)$ \\
\\[-0.35cm]
\hline
\\[-0.3cm]
~10 &0.2191(20) &0.06511(65) &0.1777(12) &0.1182(15) &0.1242(30) &0.1543(27) \\
~15 &0.2870(24) &0.05092(83) &0.2695(21) &0.0981(14) &0.2463(27) &0.1265(38) \\
~20 &0.3116(14) &0.04458(99) &0.2850(54) &0.0813(42) &0.2769(52) &0.1281(54) \\
~30 &0.3378(23) &0.0385(28)  &0.3255(98) &0.0611(74) &0.304(15)  &0.0829(97) \\
\end{tabular}
\end{center}
\end{table}


\begin{table}[ht]
\renewcommand{\tablename}{Table}
\caption{\baselineskip=0.8cm Partition function zeros for polyalanine
 chain lenghts $N$ with DDE solvent.}
\begin{center}
\begin{tabular}{lcccccc}\\
\\[-0.3cm]
$~N$~~&Re$(u_1)$ &Im$(u_1)$  & Re$(u_2)$ & Im$(u_2)$ &Re$(u_3)$  &Im$(u_3)$ \\
\\[-0.35cm]
\hline
\\[-0.3cm]
~10 &0.35256(86)&0.07316(63) &0.3329(12) &0.1424(12) &0.2959(32) &0.1907(16)\\
~15 &0.3795(19) &0.05653(89) &0.3588(21) &0.0972(15) &0.3612(87) &0.1475(71)\\
~20 &0.3794(30) &0.0487(22)  &0.3710(40) &0.0788(24) &0.3784(62) &0.1202(54)\\
~30 &0.3883(36) &0.0449(31)  &0.3976(37) &0.0774(37) &0.378(20)  &0.1143(71)\\
\end{tabular}
\end{center}
\end{table}


\begin{table}[ht]
\renewcommand{\tablename}{Table}
\caption{\baselineskip=0.8cm  Estimates of the parameter $a_2$  
         for polyalanine. The gas phase results (GP) are from 
          Ref.~19. }
\begin{center}
\begin{tabular}{lccc}\\
\\[-0.3cm]
$~N$ & GP   &  ASA   &  DDE \\
\\[-0.35cm]
\hline
\\[-0.3cm]
~10 & 1.862(46) & 1.861(18) & 1.675(23) \\
~15 & 1.664(16) & 1.750(73) & 1.70(23)  \\
~20 & 1.558(24) & 1.541(20) & 1.79(31)  \\
~30 & 1.473(30) & 2.12(19)  & 1.74(20)  
\end{tabular}
\end{center}
\end{table}


\clearpage

\begin{figure}[b]
\begin{center}
\begin{minipage}[t]{0.95\textwidth}
\centering
\includegraphics[angle=-90,width=0.72\textwidth]{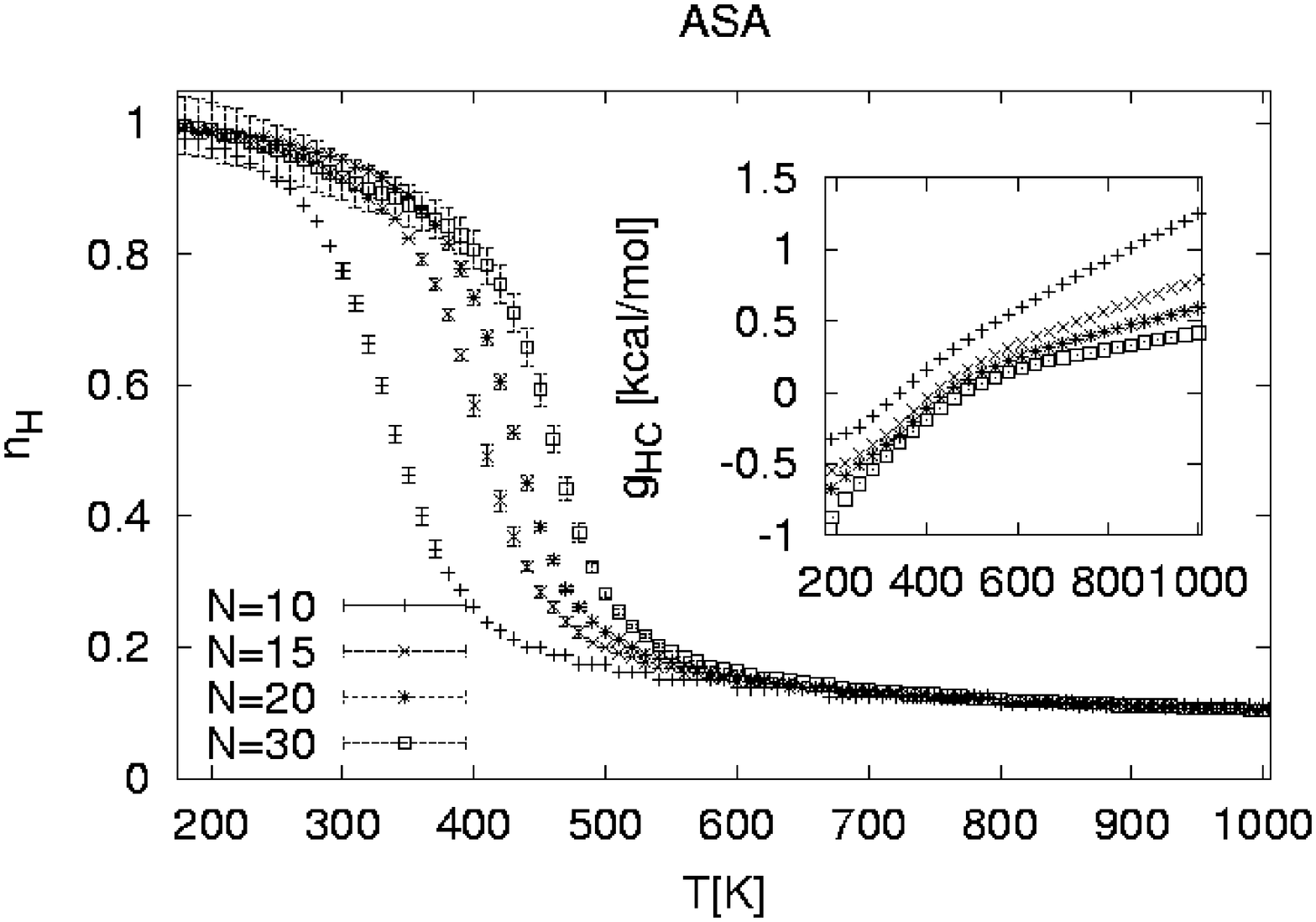}
\renewcommand{\figurename}{(Fig.1a)}
\caption{
  Temperature dependence of the  helicity order 
               parameter $q_{H} = <n_H>/(N-2)$  as obtained
               from simulations of polyalanine of chain length
               $N=10,15,20,30$ with ASA
               solvent representation. The 
               free energy difference per residue $g_{hc}$  
               between helical and coil states is shown in the
               corresponding inlets.}
\label{Fig. 1a}
\end{minipage}
\end{center}
\end{figure}

\begin{figure}[b]
\begin{center}
\begin{minipage}[t]{0.95\textwidth}
\centering
\includegraphics[angle=-90,width=0.72\textwidth]{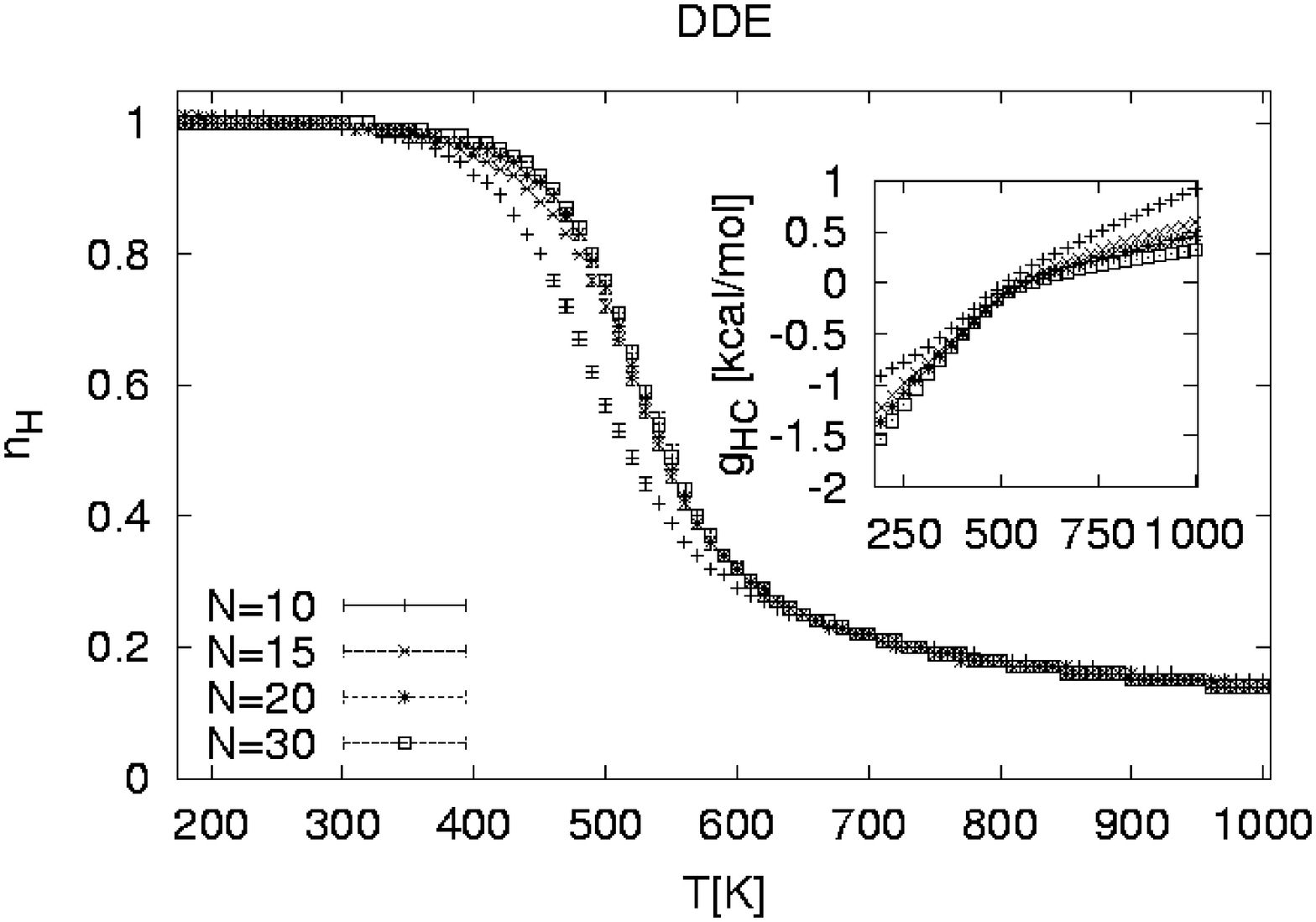}
\renewcommand{\figurename}{(Fig.1b)}
\caption{
  Temperature dependence of the  helicity order 
               parameter $q_{H} = <n_H>/(N-2)$  as obtained
               from simulations of polyalanine of chain length
               $N=10,15,20,30$ with
                DDE solvent. The 
               free energy difference per residue $g_{hc}$  
               between helical and coil states is shown in the
               corresponding inlets.}
\label{Fig. 1b}
\end{minipage}
\end{center}
\end{figure}

\newpage

\begin{figure}[b]
\begin{center}
\begin{minipage}[t]{0.95\textwidth}
\centering
\includegraphics[angle=-90,width=0.72\textwidth]{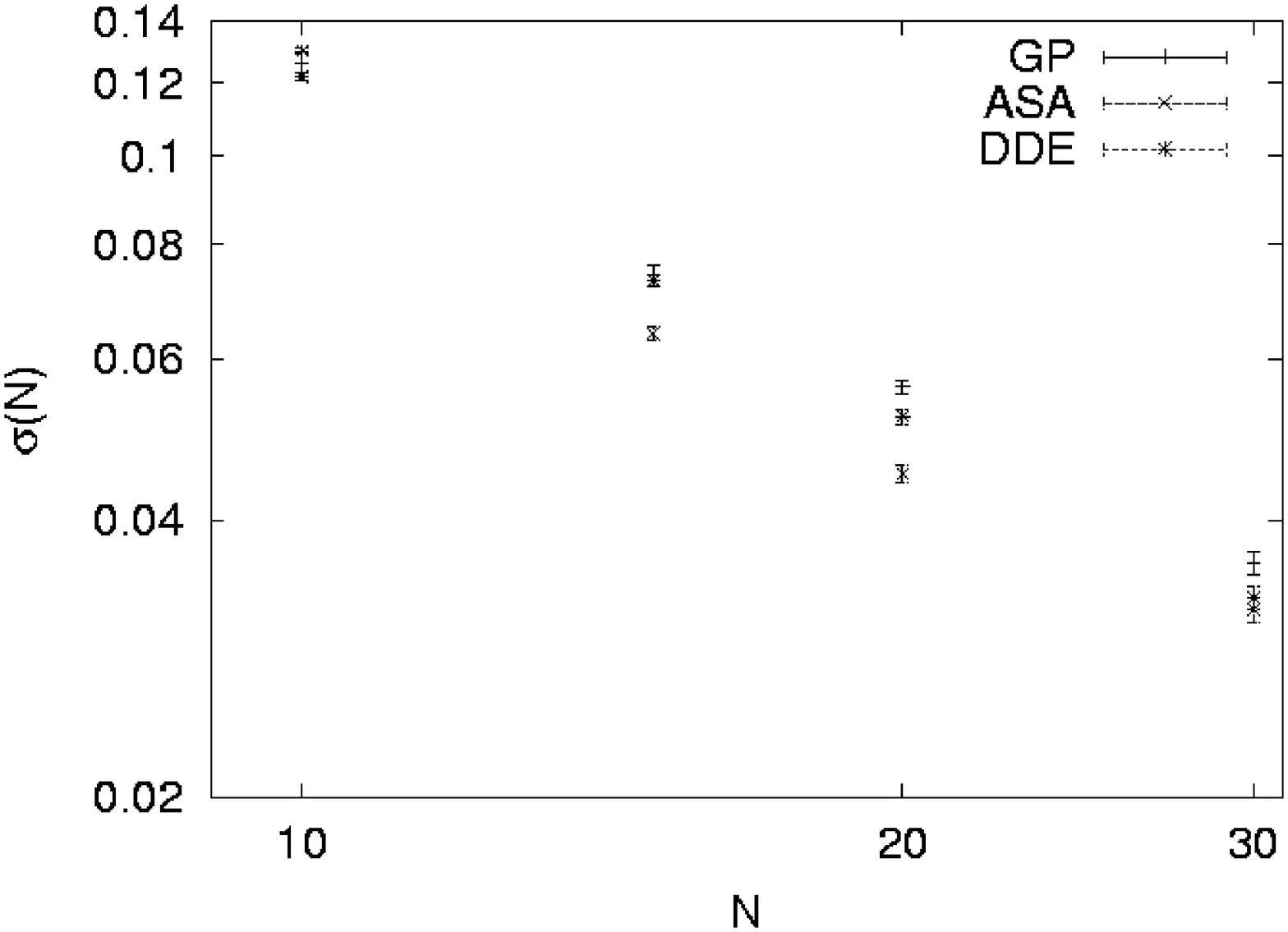}
\renewcommand{\figurename}{(Fig.2)}
\caption{
  Log-log plot of the nucleation parameter $\sigma$ as a
               function of the number of residues at $T=275$ K. The 
               gas-phase values (GP) are taken from Ref. 2.}
\label{Fig. 2}
\end{minipage}
\end{center}
\end{figure}

\begin{figure}[b]
\begin{center}
\begin{minipage}[t]{0.95\textwidth}
\centering
\includegraphics[angle=-90,width=0.72\textwidth]{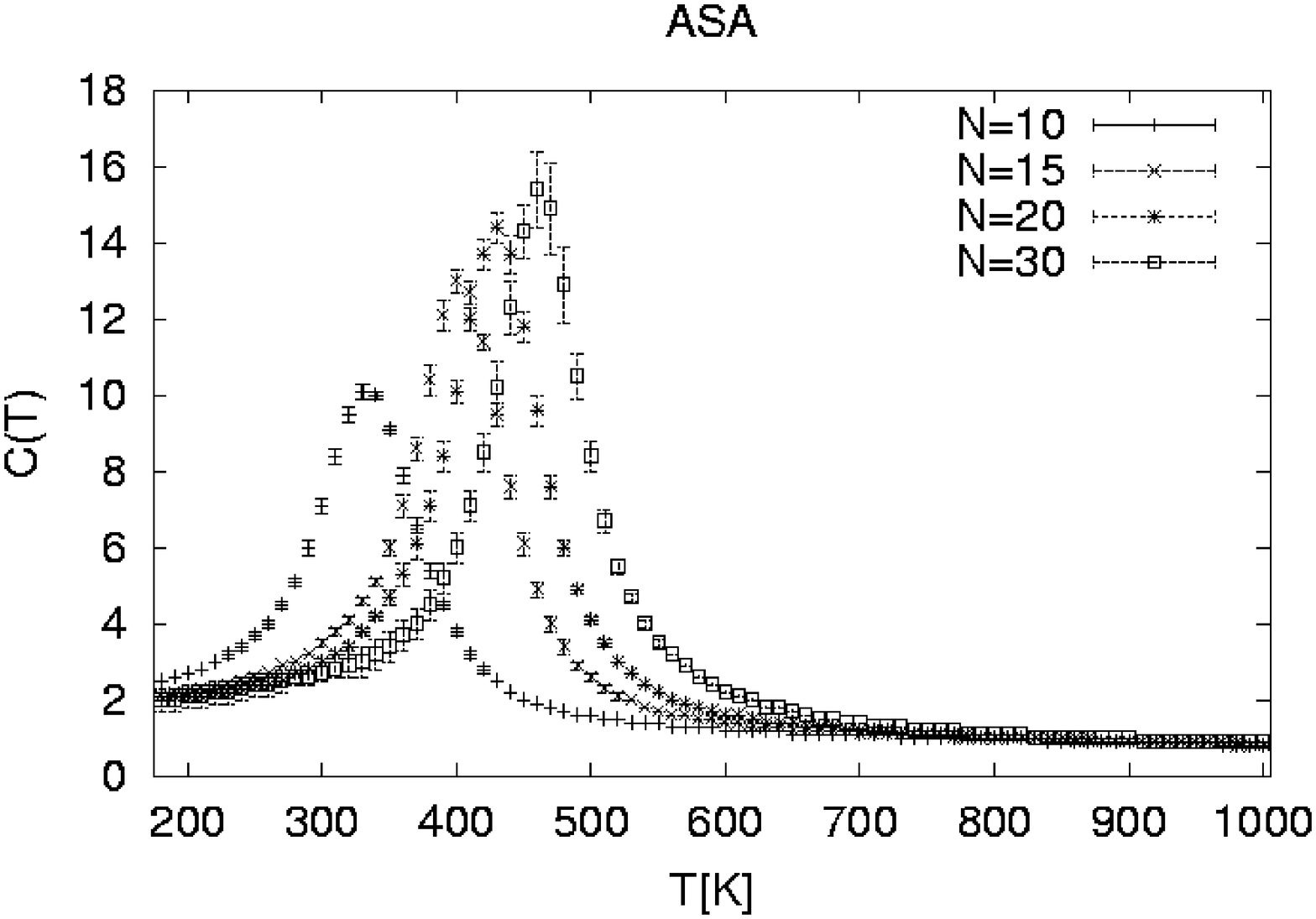}
\renewcommand{\figurename}{(Fig.3a)}
\caption{
  Specific heat $C(T)$ as a function of temperature $T$ for
               polyalanine molecules of chain length $N=10, 15, 20,$ 
               and $30$ with ASA
               solvent representation.}
\label{Fig. 3a}
\end{minipage}
\end{center}
\end{figure}

\newpage

\begin{figure}[b]
\begin{center}
\begin{minipage}[t]{0.95\textwidth}
\centering
\includegraphics[angle=-90,width=0.72\textwidth]{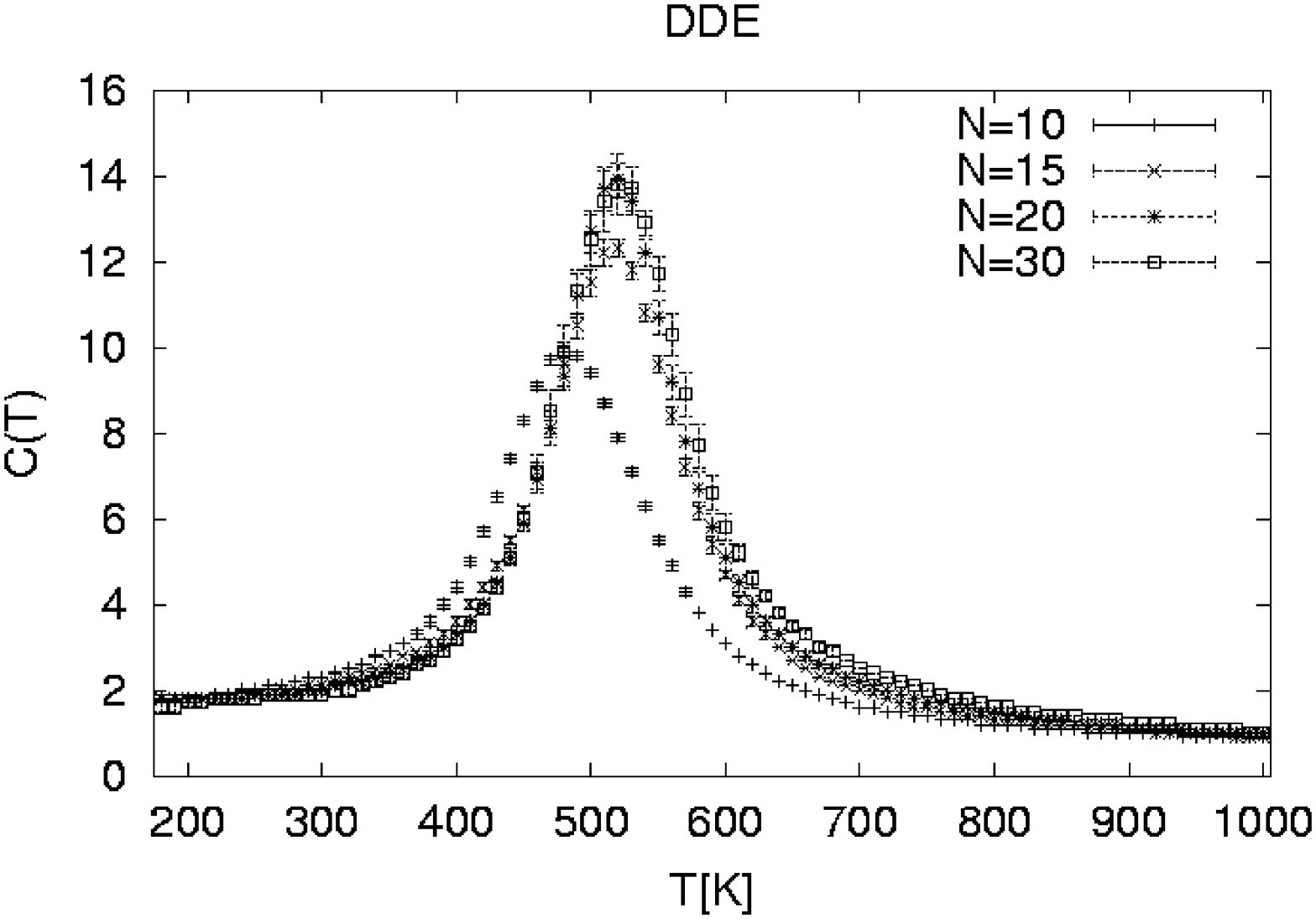}
\renewcommand{\figurename}{(Fig.3b)}
\caption{
  Specific heat $C(T)$ as a function of temperature $T$ for
               polyalanine molecules of chain length $N=10, 15, 20,$ 
               and $30$ with DDE
               solvent representation.}
\label{Fig. 3b}
\end{minipage}
\end{center}
\end{figure}

\begin{figure}[b]
\begin{center}
\begin{minipage}[t]{0.95\textwidth}
\centering
\includegraphics[angle=-90,width=0.72\textwidth]{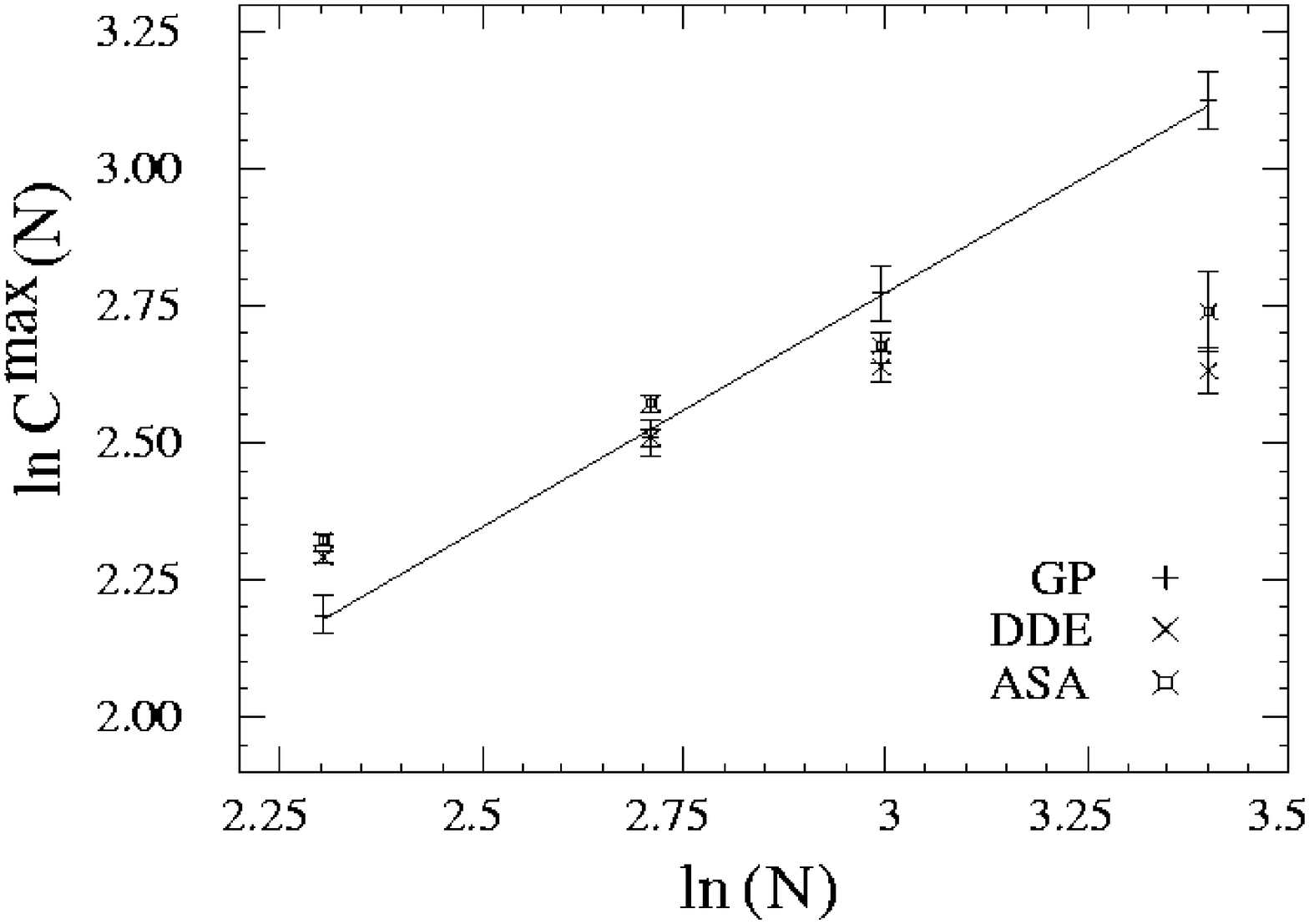}
\renewcommand{\figurename}{(Fig.4)}
\caption{
 Log-log plot of the peak value $C^{\rm max}(N)$ of the specific
               heat as a function of chain length $N$ for 
               ASA and DDE simulations. For comparison, we show also
               the corresponding gas-phase values (GP) of 
               Refs. 2,3 and the straight-line
               fit through them.}
\label{Fig. 4}
\end{minipage}
\end{center}
\end{figure}

\newpage

\begin{figure}[b]
\begin{center}
\begin{minipage}[t]{0.95\textwidth}
\centering
\includegraphics[angle=-90,width=0.72\textwidth]{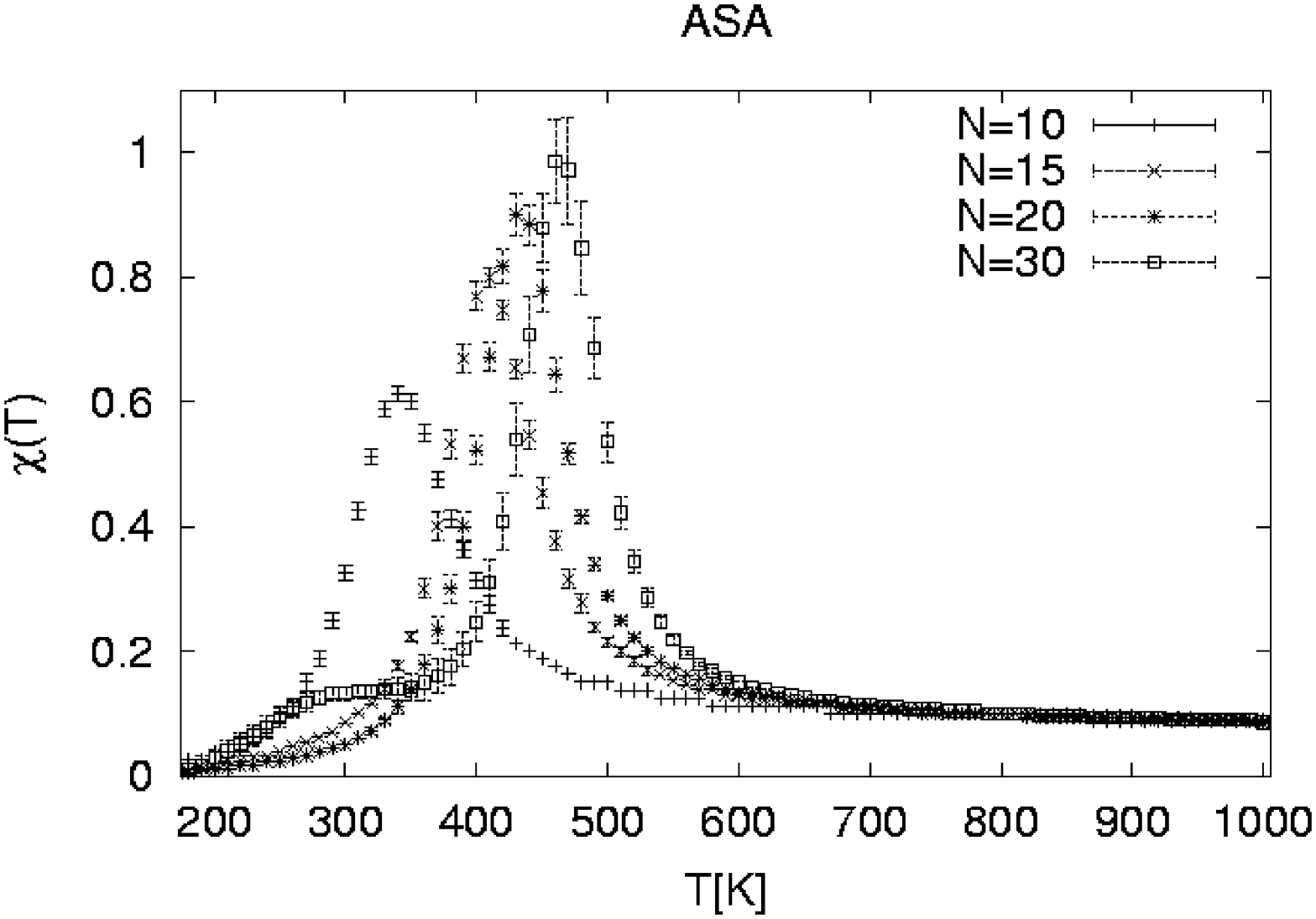}
\renewcommand{\figurename}{(Fig.5a)}
\caption{
  Susceptibility $\chi(T)$ as a function of temperature $T$ for
               polyalanine molecules of chain length $N=10, 15, 20,$ 
               and $30$ with ASA
               solvent representation.}
\label{Fig. 5a}
\end{minipage}
\end{center}
\end{figure}

\begin{figure}[b]
\begin{center}
\begin{minipage}[t]{0.95\textwidth}
\centering
\includegraphics[angle=-90,width=0.72\textwidth]{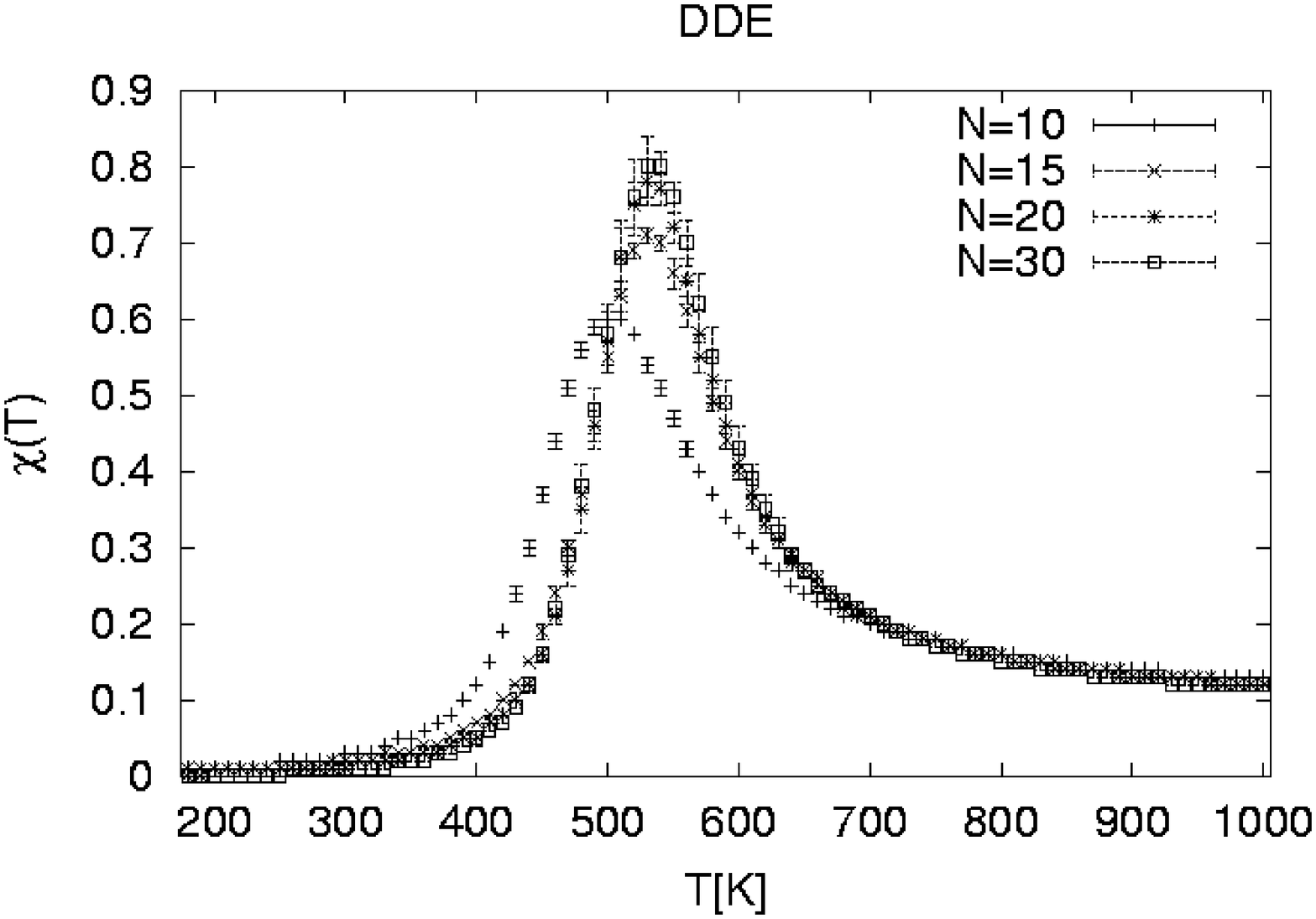}
\renewcommand{\figurename}{(Fig.5b)}
\caption{
  Susceptibility $\chi(T)$ as a function of temperature $T$ for
               polyalanine molecules of chain length $N=10, 15, 20,$ 
               and $30$ with DDE
               solvent representation.}
\label{Fig. 5b}
\end{minipage}
\end{center}
\end{figure}

\newpage

\begin{figure}[b]
\begin{center}
\begin{minipage}[t]{0.95\textwidth}
\centering
\includegraphics[angle=-90,width=0.72\textwidth]{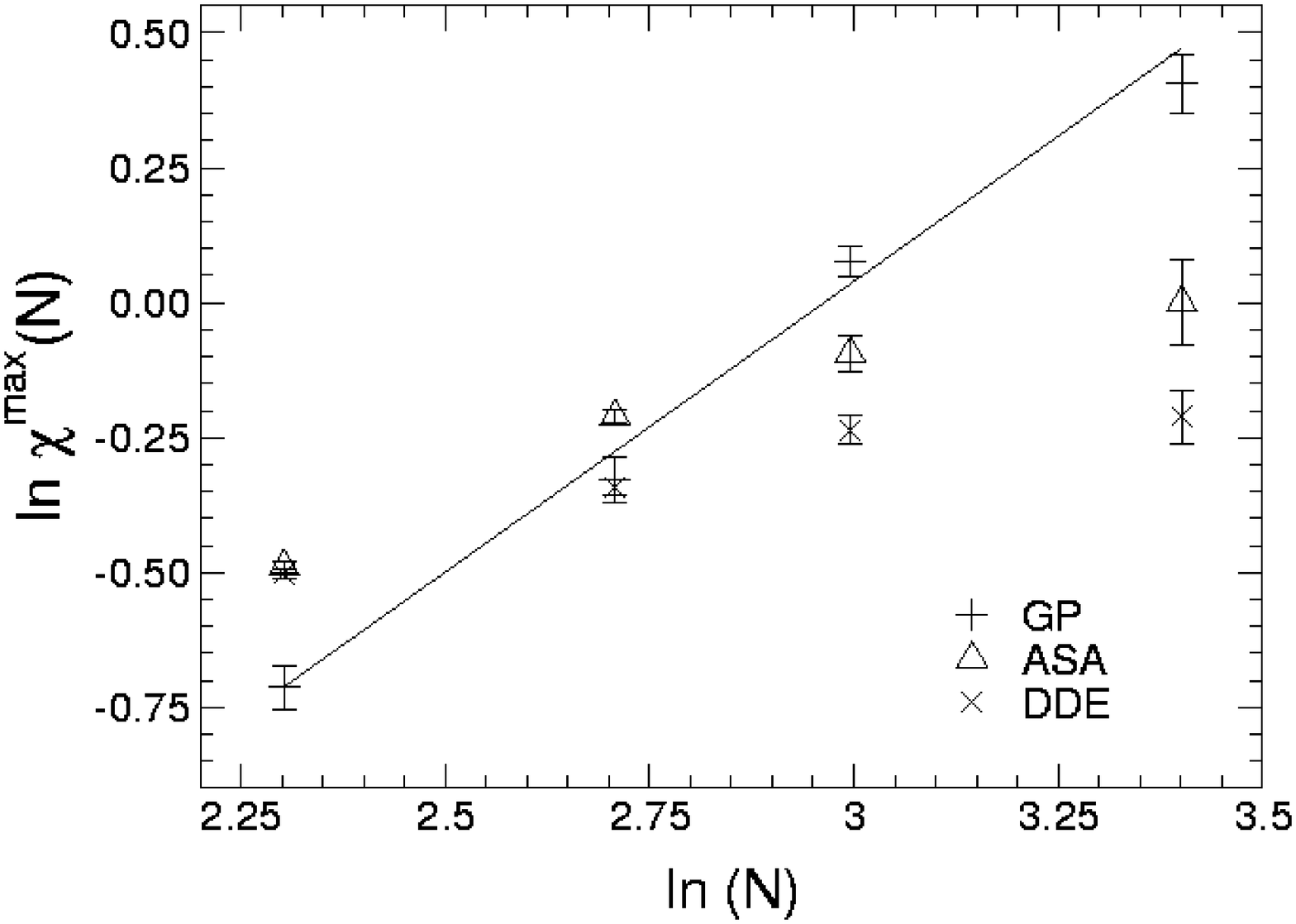}
\renewcommand{\figurename}{(Fig.6)}
\caption{
  Log-log plot of the peak value $\chi^{\rm max}(N)$ of the 
               susceptibility as a function of chain length $N$ for 
               ASA  and DDE simulations. The corresponding gas-phase 
               values (GP) of Ref. 2 and the straight-line
               fit through them are shown for comparison.}
\label{Fig. 6}
\end{minipage}
\end{center}
\end{figure}

\begin{figure}[b]
\begin{center}
\begin{minipage}[t]{0.95\textwidth}
\centering
\includegraphics[angle=-90,width=0.72\textwidth]{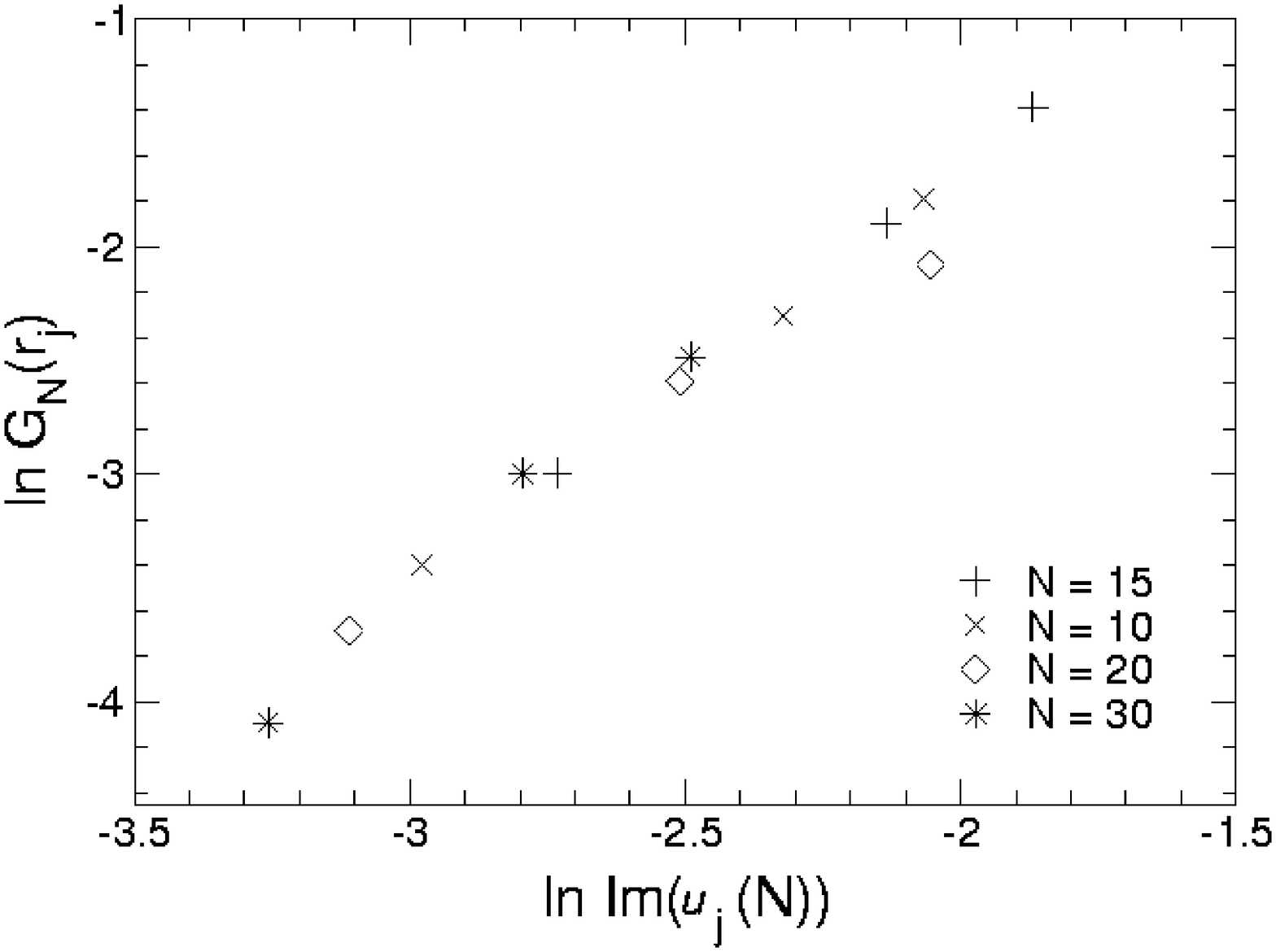}
\renewcommand{\figurename}{(Fig.7)}
\caption{
 Distribution of zeros for the polyalanine model with
               ASA solvent. }
\label{Fig. 7}
\end{minipage}
\end{center}
\end{figure}

\newpage

\begin{figure}[b]
\begin{center}
\begin{minipage}[t]{0.95\textwidth}
\centering
\includegraphics[angle=-90,width=0.72\textwidth]{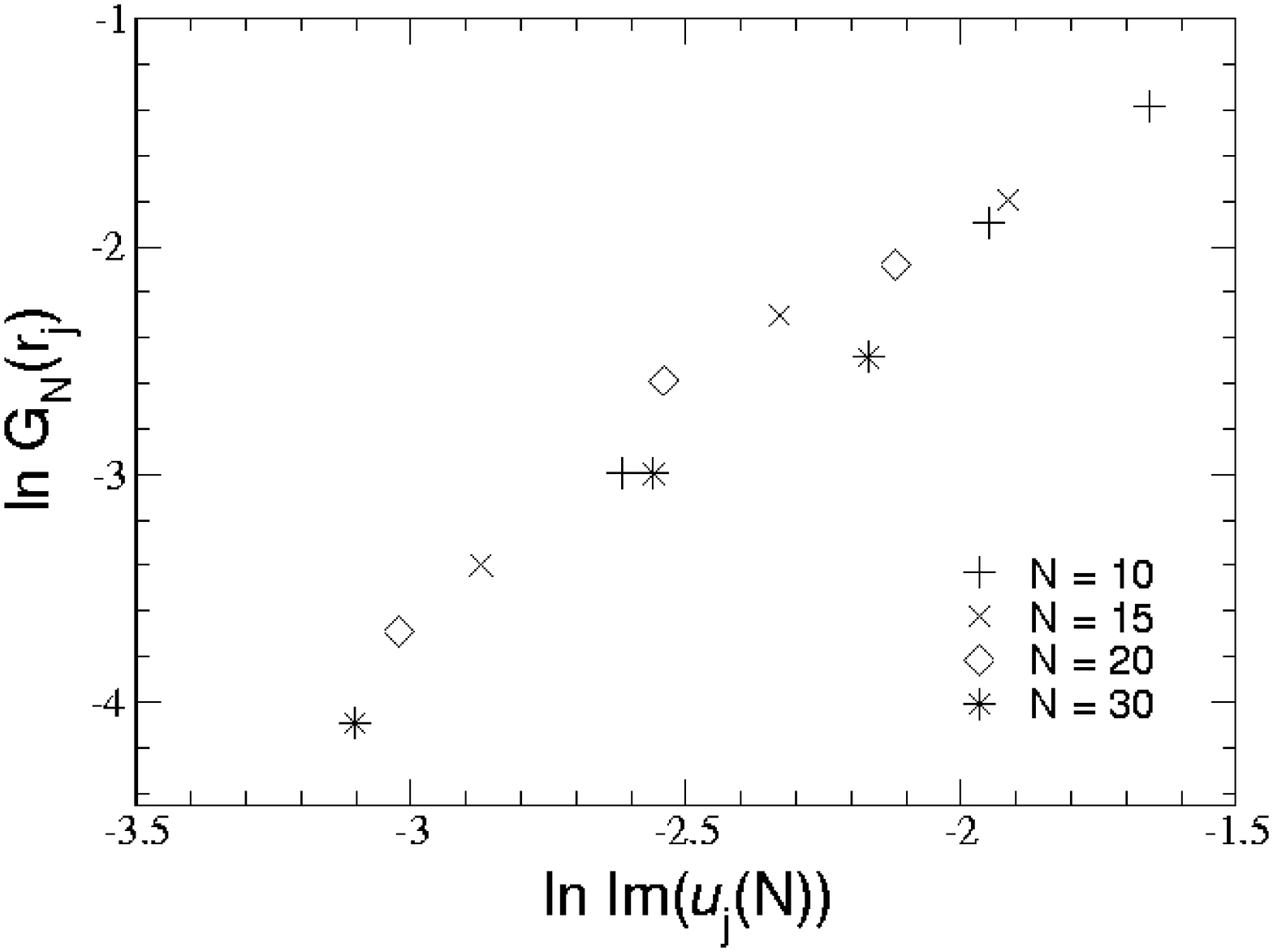}
\renewcommand{\figurename}{(Fig.8)}
\caption{
 Distribution of zeros for the polyalanine model with
               DDE solvent. }
\label{Fig. 8}
\end{minipage}
\end{center}
\end{figure}


\begin{references}
\bibitem{Poland} D.~Poland and H.A.~Scheraga, {\it Theory of Helix-Coil
         Transitions in Biopolymers} (Academic Press, New York, 1970).
\bibitem{HO98c} U.H.E.~Hansmann and Y.~Okamoto, {J. Chem.~Phys.} {\bf 110},
                1267 (1999); {\bf 111} 1339(E) (1999).
\bibitem{AH99b} N.A. Alves  and U.H.E. Hansmann,\
                {Phys. Rev. Lett.}  {\bf 84}, 1836 (2000).
\bibitem{AH00b} N.A. Alves and U.H.E.~Hansmann, {Physica A}
                {\bf 292}, 509 (2001).
\bibitem{Jarrold} R.R. Hudgins, M.A. Ratner and M.F. Jarrold,
                  {J. Am. Chem. Soc.} {\bf 120}, 12974 (1998).
\bibitem{MO1}  A. Mitsutake and Y. Okamoto, {J. Chem. Phys.} {\bf 112},
               10638 (2000).
\bibitem{PH01h} Y.~Peng and U.H.E.~Hansmann, {Biophys. J.}
                {\bf 82}, 3269 (2002). 
\bibitem{oons} T. Ooi, M. Obatake, G. Nemethy, H.A. Scheraga,
               {Proc. Natl. Acad. Sci.  USA} {\bf 8}, 3086 (1987).
\bibitem{lavarty} R. Lavery, H. Sklenar, K. Zakrzewska, B. Pullman,
                  {J. Biomol. Struct. \& Dynamics} {\bf  3}, 989 (1986).
\bibitem{EC} M.J. Sippl, G. N{\'e}methy, and H.A. Scheraga,
             {J. Phys. Chem.} {\bf 88}, 6231~(1984), 
             and references therein.
\bibitem{SMMP} F.~Eisenmenger, U.H.E.~Hansmann, Sh.~Hayryan, C.-K.~Hu,
               {Comp.~Phys.~Comm.} {\bf 138}, 192 (2001).
\bibitem{hingerty} B. Hingerty, R.H. Richie, T.L. Ferrel, J. Turner,
                  {Biopolymers} {\bf 24}, 427 (1985).
\bibitem{Review} U.H.E.~Hansmann and Y.~Okamoto, 
         in: Stauffer, D. (ed.) ``{\it Annual Reviews in Computational 
         Physics VI}'',(Singapore: World Scientific), p.129.  (1998).
\bibitem{HO} U.H.E. Hansmann and Y. Okamoto,  J.~Comp.~Chem.
             {\bf 14}, 1333 (1993).
\bibitem{MU} B.A. Berg  and T. Neuhaus,
             {Phys. Lett.} B {\bf 267}, 249 (1991).
\bibitem{OH95b} Y.~Okamoto and U.H.E.~Hansmann,  J.~Phys.~Chem.
                {\bf 99}, 11276 (1995).
\bibitem{FS} A.M. Ferrenberg and R.H. Swendsen, {Phys.\ Rev.\ Lett.}
             {\bf  61}, 2635 (1988); {Phys. Rev. Lett.} {\bf 63},
             1658(E) (1989), and
              references given in the erratum.
\bibitem{Berg}  B.A. Berg, {J. Stat. Phys.} {\bf 82}, 323 (1996).
\bibitem{AFH01d} N.A. Alves, J.P.N. Ferrite and U.H.E. Hansmann,
                {Phys. Rev. E} {\bf 65}, 036110 (2002).
\bibitem{fisher} M.E. Fisher, in {\it Lectures in Theoretical Physics},
                 vol. 7c p.1 (University of Colorado Press, Boulder, 1965).
\bibitem{itzykson} C. Itzykson, R.B. Pearson, and J.B. Zuber,
                   Nucl. Phys. B {\bf 220} [FS8], 415 (1983).
\bibitem{Wales} P.N. Mortenson, D.A. Evans and D.J. Wales, 
                {J. Chem. Phys.} {\bf 117}, 1363 (2002). 
\bibitem{VRS00} J.A. Vila, D.R. Ripoll and H.A. Scheraga, 
                Proc. Natl. Acad. Sci. USA {\bf 97}, 13075 (2000).
\bibitem{LJB01} Y. Levy, J. Jortner and O.M. Becker, 
                Proc. Natl. Acad. Sci. USA {\bf 98}, 2188 (2001).
\bibitem{ZB} B.H. Zimm and J.K. Bragg,  J. Chem. Phys. {\bf 31}, 526 (1959).
\bibitem{CB} Chakrabartty, A.; R.L. Baldwin, R.L. In
    {\it Protein Folding: In Vivo and In Vitro}; Cleland J.;
    King, J. eds.; ACS Press: Washington, D.C., 1993; pp. 166--177.
\bibitem{ADH97} N.A. Alves, J.R.~Drugowich de Felicio and U.H.E.~Hansmann,
                {Int.~J.~Mod.~Phys.~C} {\bf 8}, 1063 (1997).
\bibitem{AHP01e} N.A. Alves, U.H.E. Hansmann and Y. Peng,
                 {Int. J. Mol. Sci.} {\bf 3}, 17 (2002).
\bibitem{JK}   W. Janke and R. Kenna, J. Stat. Phys. {\bf 102}, 1211 (2001); 
               Nucl. Phys. Proc. Supl. {\bf B106}, 905 (2002); 
               {\it http://arXiv.org/abs/cond-mat/0103333};
               {\it http://arXiv.org/abs/cond-mat/0208014}.
\end{references}
\end{document}